\documentclass[aps,prb,showpacs,superscriptaddress,twocolumn,amsmath,amssymb]{revtex4-2}
\usepackage{graphicx,bm,psfrag}
\usepackage{color,xfrac,xcolor}
\usepackage{braket}
\usepackage[hidelinks , colorlinks = true, citecolor = blue, urlcolor = blue, linkcolor = blue]{hyperref}

\newcommand{\figref}[1]{Fig. \ref{#1}}

\begin{document}

    \title{Current noise in quantum dot thermoelectric engines}
	\author{Simon Wozny}
	\email{simon.wozny@ftf.lth.se}
	\affiliation{NanoLund and Solid State Physics, Lund University, Box 118, 22100 Lund, Sweden}
	\author{Martin Leijnse}
	\affiliation{NanoLund and Solid State Physics, Lund University, Box 118, 22100 Lund, Sweden}
	
	\date{\today}
	
\begin{abstract}
    We theoretically investigate a thermoelectric heat engine based on a single-level quantum dot, calculating average quantities such as current, heat current, output power, and efficiency, as well as fluctuations (noise).
    Our theory is based on a diagrammatic expansion of the memory kernel together with counting statistics, and we investigate the effects of strong interactions and next-to-leading order tunneling.
    Accounting for next-to-leading order tunneling is crucial for a correct description when operating at high power and high efficiency, and in particular affect the qualitative behavior of the Fano factor and efficiency.
    We compare our results with the so-called thermodynamic uncertainty relations, which provide a lower bound on the fluctuations for a given efficiency.
    In principle, the conventional thermodynamic uncertainty relations can be violated by the non-Markovian quantum effects originating from next-to-leading order tunneling, providing a type of quantum advantage.
    However, for the specific heat engine realization we consider here, we find that next-to-leading order tunneling does not lead to such violations, but in fact always pushes the results further away from the bound set by the thermodynamic uncertainty relations.
\end{abstract}

\maketitle

\section{Introduction}

Transport through quantum dot (QD) systems has been studied intensely in recent years.
They can be used to investigate fundamental aspects of open quantum systems, non-equilibrium and many body effects \cite{Kouwenhoven1997, Cronenwett1998, Sasaki2000, Wiel2002, Su2016}.
In addition, QDs are used in many quantum technological applications, e.g. sensors \cite{Hu2007,Kiyama2018,Arquer2021} and qubits \cite{Loss1998,Chatterjee2021,Burkard2023}.

From a theory perspective, transport in QDs (and other interacting nanoscale systems) presents an interesting challenge.
In the absence of interactions, scattering theory can provide exact results for many simple systems, including a single resonant level \cite{Blanter2000}.
In the presence of electron-electron interactions this problem gets significantly more complicated and cannot be solved exactly.
At low temperatures, non-equilibrium Green's function methods have been used together with renormalization group approaches \cite{Rammer2007,Metzner2012}, while others used fully numerical methods \cite{Rubtsov2005,Schollwoeck2005,Anders2008,Jin2008,Weiss2008,Werner2009}.

Master equations naturally include Coulomb interactions, at the cost of treating the tunneling within some level of approximation.
In leading order perturbation theory only sequential tunneling is considered.
There are also perturbative schemes that allow including higher order.
In Ref. \cite{Averin1990} elastic and inelastic co-tunneling have been considered, as it is the dominating contribution in the Coulomb blockade regime, where the sequential tunneling contributions to the current are exponentially suppressed.
Later, all next to leading order processes \cite{Koenig1997, Leijnse2008, Koller2010} and more general higher order processes \cite{Schoeller1994, Pedersen2005, Saptsov2012} were included.

\begin{figure}
    \centering
    \includegraphics[width=\columnwidth]{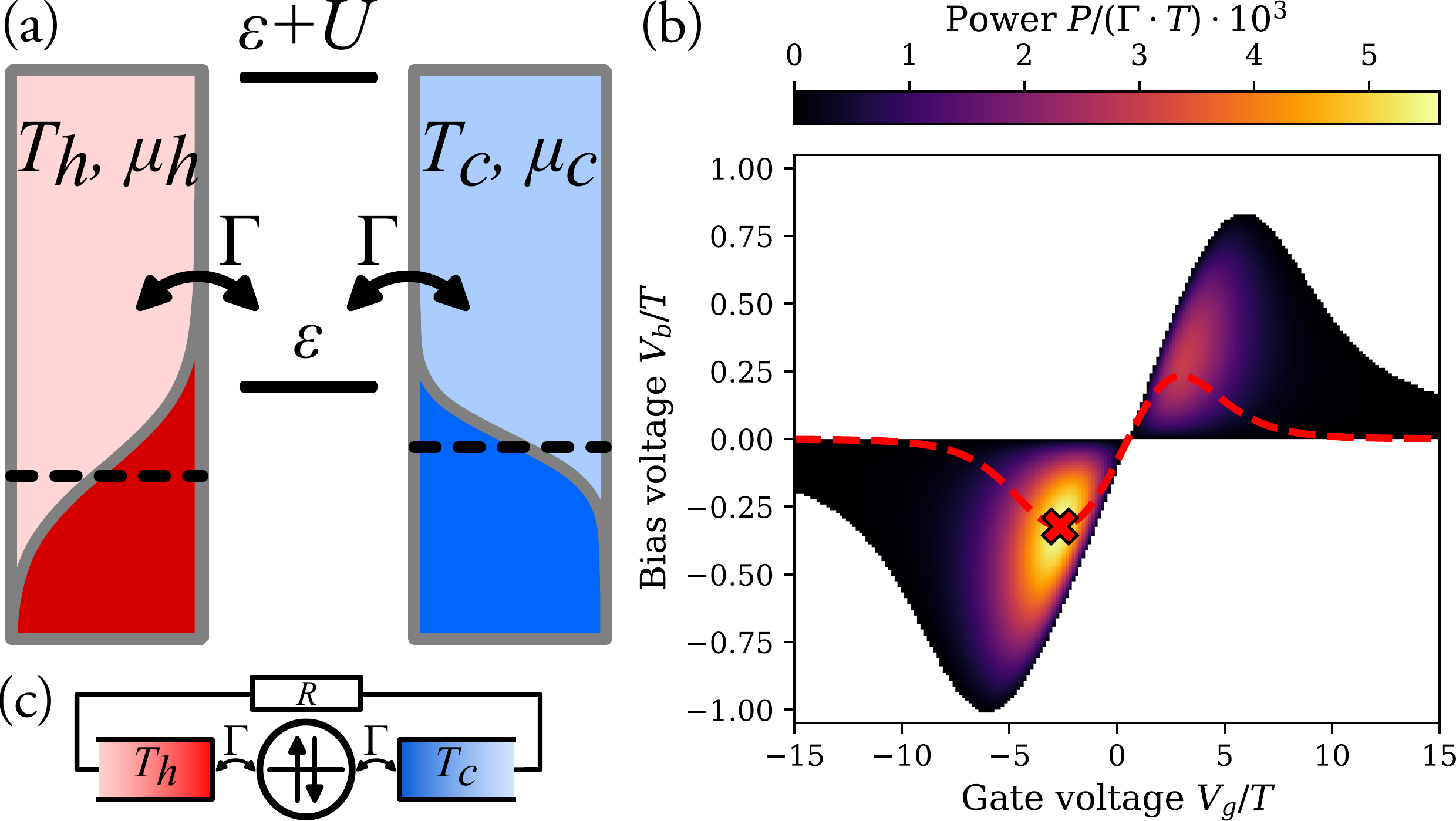}
    \caption{
    (a) Sketch of the system with the QD energy level $\varepsilon$ coupled to a hot (h) and cold (c) lead described by temperatures $T_{h/c}$, chemical potentials $\mu_{h/c}$ and tunnel couplings $\Gamma$.
    (b) Power from a gate/bias sweep with temperatures $T_h=1.3$, $T_=1.0$, $U=100T_c$ and $\Gamma=0.25 T_c$ at symmetric bias. Note the small bias/gate range, focusing on a region close to the $0 \-- 1$ charge degeneracy point. The red curve is the load voltage depending on the gate voltage with a resistance coupled according to (c). The load resistance $R\approx 74.43$ is chosen such that the cut goes through the maximum power point.
    (c) Coupling scheme for a load resistor.
    }
    \label{fig:1}
\end{figure}

High tunability and control allows QD devices to be operated as thermal machines in many different forms.
They can be used as heat pumps or heat valves \cite{Mitchison2019,Dutta2020}, for heat rectification \cite{Malik2022,Tesser2022} and refrigeration \cite{Edwards1993,Mitchison2019}.
There are also many schemes in which the QDs are operated as heat engines.
Quantum mechanical equivalents of Otto or Carnot cycles \cite{Alicki1979,Bhandari2020,Dann2020} make use of cyclic driving of a QD between two thermal baths.
In  engines based on coupling to a single heat bath there are autonomous Maxwell demon type engines \cite{Josefsson2020a,Josefsson2020, AnnbyAndersson2024} as well as Szilard engine realizations \cite{Koski2014,Parrondo2015,Barker2022,Barker2022Thesis}, where measurement and feedback play an important role.

In contrast to cyclic schemes, steady state thermoelectric heat engines use the energy filtering capabilities of the quantized QD levels to extract electrical work from a temperature gradient between leads \cite{Humphrey2002,Humphrey2005,O’Dwyer2006,Esposito2009,Kennes2013,Benenti2017}.
A QD (that can be modeled by a single spinful level) coupled to metallic leads has been investigated earlier and it has been shown that higher order tunneling contributions play an important role in limiting the efficiency \cite{Josefsson2018,Josefsson2019}.

In addition to the average current, it is interesting to study the current fluctuations or noise, which can contain additional information about the system \cite{Hershfield1993,Thielmann2003}. 
Noise in charge transport through QDs has been investigated theoretically \cite{Hershfield1993,Sukhorukov2001,Thielmann2003,Thielmann2005,Thielmann2005a,Aghassi2008,Kaasbjerg2015} and experimentally \cite{Gustavsson2007,Hasler2015}.
Co-tunneling has been shown to play an important role and can alter the noise characteristics \cite{Sukhorukov2001,Thielmann2005,Thielmann2005a}.
Full counting statistics can be used to calculate the noise and  higher current cumulants \cite{Flindt2008,Flindt2010,Emary2009}.

The noise in QD systems plays an important role in the context of fluctuation dissipation relations \cite{Sukhorukov2001, Tesser2024} and thermodynamic uncertainty relations (TURs) \cite{Barato2015,Pietzonka2018,Seifert2018,Gingrich2016,Agarwalla2018,Liu2019,Kheradsoud2019}.
The TUR is a lower bound on the fluctuations of the current (and therefore output power) through the heat engine, that increases with increasing efficiency and current.
Alternatively, it provides an upper bound on the efficiency that decreases with decreasing fluctuations.
It is known that the classically derived TUR holds in the sequential tunneling regime, but quantum mechanical effects and higher order tunneling contributions can lead to violations \cite{Ptaszyifmmodenelsenfiski2018,Agarwalla2018,Liu2019}, giving a type of quantum advantage.

In this work, we investigate the current noise together with the heat current and generated electric power in a single spinful level QD operated as a thermoelectric engine, including tunnel processes to next-to-leading order.
We calculate these quantities for a given set of parameters in a realistic heat engine setup, varying the QD level and comparing with sequential tunneling and non-interacting results.
Similarly, we calculate these quantities at the maximum power point, varying the tunnel coupling and temperature difference, in order to systematically investigate the effects of interactions and next-to-leading order tunneling.

The paper is organized as follows.
In section II we introduce the setup and briefly review the Master equation and counting statistics approach we use.
Section III investigates the influence of interactions as well as the next-to-leading order contributions on the noise in the thermoelectric engine setup and section IV summaries the results and provides an outlook.
Throughout the paper we use units where $\hbar = e = k_B = 1$.

\section{Model and Methods}

We investigate a QD described by a single spinful level coupled to a hot (h) and a cold (c) lead, schematically shown in \figref{fig:1}(a).
The QD is described by the level $\varepsilon$ and the Coulomb charging energy is given by $U$.
With the QD creation/annihilation operators $d^\dagger_{\sigma}/d_{\sigma}$ the QD Hamiltonian reads
\begin{align}
    H_D =& \sum_{\sigma=\uparrow,\downarrow} n_\sigma \varepsilon_\sigma + U n_\uparrow n_\downarrow, \qquad n_\sigma=d^\dagger_\sigma d_{\sigma}.
\end{align}
We assume infinite and non-interacting fermionic leads described by their chemical potentials $\mu_{r}$ and temperatures $T_{r}$, where $r=h/c$.
With the lead creation/annihilation operators $c^\dagger_{k\sigma r}/c_{k\sigma r}$, where $k$ is a momentum index, $\sigma$ is the spin index and $r$ the lead index, the lead Hamiltonian can be written as
\begin{align}
    H_r =& \sum_{k,\sigma} \omega_{k\sigma l} n_{k \sigma l}, \qquad n_{k \sigma r}=c^\dagger_{k\sigma r}c_{k\sigma r}.
\end{align}
The leads and the QD are coupled by the tunneling amplitudes $t_{k\sigma r}$ via the tunneling Hamiltonian 
\begin{align}
    H_{T,r} =& \sum_{k,\sigma} t_{k \sigma r} d^\dagger_{\sigma}c_{k\sigma r} + \mathrm{H.c.}.
\end{align}
The tunnel amplitude defines the bare tunnel rates
\begin{align}
    \Gamma_r = 2\pi \nu_r |t_{k\sigma r}|^2,
\end{align}
with the lead density of states $\nu_r$.
We assume the wide band limit, where $\nu_r$ is constant over an energy range much larger than all other relevant energies.
Combining these parts gives the full Hamiltonian
\begin{align}
    H =& H_D + \sum_{r} H_{T,r} + \sum_{r} H_{r}.
\end{align}

\subsection{Transport quantities and counting statistics}

To calculate the various transport quantities for this system we use the real-time diagrammatic technique in Liouville-Laplace superoperator space 
\cite{Schoeller1994, Koenig1997, Leijnse2008, Koller2010, Saptsov2012}.
For the practical implementation see
\footnote{
To perform the calculations of the transport quantities we use a custom version of the QD transport package QmeQ \cite{Kirifmmodeselsesfianskas2017}.
The standard version can perform current and heat current calculations and has been extended to include calculations of the current noise.
The extended version used here is available at \url{https://doi.org/10.5281/zenodo.14046688}.
}.
Starting from the Liouville-von Neumann equation for the density matrix of the full system
\begin{align}
    \dot{\rho}(t) = -i[H,\rho (t)] = \mathcal{L}\rho (t) 
\end{align}
we perform a Laplace transformation and trace over the reservoirs.
The stationary state reduced density matrix of the QD, $\rho_D$, can then be found via the Laplace zero frequency limit and we need to solve
\begin{align}
    (\mathcal{L}_D + W(i0^+))\rho_D = 0, \quad \mathcal{L}_D\bullet = -i[H_D,\bullet].
\end{align}
This expression is in principle exact, but in practice some form of approximation is needed to find $W(i0^+)$.
We expand in the tunneling Hamiltonian $H_T$ up to next-to-leading order.
Here leading / next-to-leading order is $H_T^2 / H_T^4$,  which is $\Gamma / \Gamma^2$.
We count orders in $\Gamma$ and refer to it as first and second order perturbation theory.
Details on this derivation and expressions for the first and second order kernels can be found Refs \cite{Leijnse2008,Emary2009}.

In order to calculate the noise (second current cumulant) we find the counting field resolved kernel $W(z,\chi_r)$ where $z$ is the Laplace frequency and $\chi_r$ is the counting field at lead $r$.
As described in Ref. \cite{Emary2009} this can be done by modifying the bath contraction functions in the integrals that make up the kernel to include the counting field.
In second order the counting field resolved kernel can be expanded as
\begin{align}
    W(\chi_r,z) = \sum_{k=-2,...,2} e^{ik \chi_r} W_{k,r}(z).
\end{align}
The kernels $W_{k,r}$ then contain all tunneling processes that transfer $k$ electrons to/from lead $r$.
The sub-kernels $W_{\pm 2,r}$ contain exclusively second-order terms, while the others can contain both first and second-order contributions.
In a two-terminal setup the current and noise at both lead are the same, due to current conservation, and we therefore drop the lead indices on these quantities.

To calculate the current cumulants we are interested in we follow Refs \cite{Flindt2008,Flindt2010,Emary2009}.
We use $|\bullet)$ to denote vectors in Liouville-Laplace superoperator space.
In this notation expectation values are given by 
\begin{align}
    (( \bullet )) = (\psi_0 | \bullet | \psi_0) = \mathrm{Tr}(\bullet \rho_D),
\end{align}
where the left and right null vectors of $W(0,i0^+)$ are defined via
\begin{align}
    W(0,i0^+)|\psi_0)=(\psi_0|W(0,i0^+)=0.
\end{align}
We define projection operators in super operator space
\begin{align}
    \mathcal{P}=|\psi_0)(\psi_0|, \quad \mathcal{Q}=1-\mathcal{P}
\end{align}
and the pseudo-inverse
\begin{align}
    \mathcal{R}(\delta) = \mathcal{Q} \frac{1}{\delta+W(0,\delta+i0^+)} \mathcal{Q}.
\end{align}
The first two current cumulants, i.e. mean particle current $I$ and current noise $S$ then read \cite{Flindt2008,Emary2009,thesisFlindt}
\begin{align}
    I =& - (( J' ))\\
    S =& - ((J'' - 2 J' \mathcal{R} J')) + 2 I ((\dot{J}' - J' \mathcal{R} \dot{J})),
\end{align}
where  $\mathcal{R}=\mathcal{R}(\delta\rightarrow 0)$ and $J(\chi,\delta)$ is the shifted kernel
\begin{align}
    J(\chi,\delta) = W(\chi,z=\delta-i0^+)-W(\chi=0,z=i0^+),
\end{align}
with the derivatives
\begin{align}
    J' = \partial_\chi J|_{\chi,\delta\rightarrow 0}, \quad \dot{J} = \partial_\delta J|_{\chi,\delta\rightarrow 0}.
\end{align}
From the current $I$ and the noise $S$ the Fano-factor can be calculated as
\begin{align}
    F = \frac{S}{|I|}.
\end{align}

The heat current $Q_r$ out of lead $r$ can be calculated from the charge current $I_r$ and energy current $J_{E,r}$ in that lead via $Q_r=J_{E,r}-\mu_r I_r$.
Note that unlike the charge and energy currents, the heat current is not conserved and we need to keep the reservoir index.
Details on the calculations of the energy current can be found in Refs \cite{Saptsov2012,Josefsson2019,Josefsson2020}.

In the case of zero Coulomb interaction $U=0$ the equations for the different spins decouple and the system can be solved analytically using Landauer-B\"uttiker scattering theory \cite{Blanter2000}.
The resulting equations for the transport quantities of interest can be found in Appendix \ref{app:lb}.

\subsection{Operation as thermoelectric engine}

The QD system can be operated in a parameter regime where a temperature bias can be utilized to perform electrical work, see Refs. \cite{Humphrey2002,O’Dwyer2006,Esposito2009,Kennes2013,Benenti2017,Josefsson2018,Josefsson2019}.
The QD level acts as an energy filter, causing electrons to flow from the lead with higher to the lead with lower occupation at energy $\varepsilon$.
In the situation depicted in \figref{fig:1}(a) $\varepsilon$ lies above the chemical potentials.
The chemical potential of the hot lead $\mu_h$ lies below that of the cold lead $\mu_c$, i.e. there is an electric bias from cold to hot lead.
However, due to the broader Fermi function in the hot lead the occupation at energy $\varepsilon$ in the hot lead can be higher than in the cold lead leading to a current against that bias.

Important quantities characterizing thermoelectric engines are the (electrical) power $P$ and the efficiency $\eta$.
In \figref{fig:1}(b) the power of a single QD heat engine is shown in gate-bias space with fixed temperature bias, tunnel couplings and charging energy.
The gate voltage $V_g$ here corresponds directly to the QD levels via $\varepsilon=-V_g$ (for simplicity setting the charge offset and gate lever arm to unity), while the bias voltage is applied symmetrically to the leads, i.e. $\mu_h=-\mu_c=\frac{V_b}{2}$.
The power due to the transport of electrons against that bias is calculated from the current and bias as
\begin{align}
    P = - I V_b.
\end{align}
Note that \figref{fig:1}(b) only show a small region around the $0-1$ charge degeneracy point where the output power is positive.
In the white areas, the QD does not operate as a heat engine.

To extract electrical work from the heat engine one cannot apply an external bias voltage.
Instead, the external load can be modeled by a resistor $R$ as shown in \figref{fig:1}(c).
The current produced by the thermally biased QD passes through the resistor and produces power there.
Due to current conservation, for every gate voltage we can find the corresponding voltage $V$ that develops over the resistor by self consistently solving \cite{Josefsson2018,Josefsson2019}
\begin{align}
    I(V_g,V)=V/R.
\end{align}
The load resistance $R$ thus leads to a built up voltage $V$ over the QD, see the dashed line in \figref{fig:1}(b).
In this sense operating the QD at a given gate and bias is equivalent to operating it as a heat engine at that gate with a specific load.
$R$ can be optimized for the operation point with e.g. maximal power output or maximal efficiency.
In this work we choose to optimize for the power $P$, marked with a cross in \figref{fig:1}(b), for detail see Appendix \ref{app:maxpp}.
In the 0 electron sector $V$ is negative, while to the right of the charge degeneracy point $V$ is positive.

The efficiency of the engine is the ratio of power and heat current at the hot lead, i.e.
\begin{align}
    \eta = \frac{P}{Q_h} = \frac{-IV}{Q_h}
\end{align}
A fundamental thermodynamic bound on the efficiency is the Carnot-efficiency 
\begin{align}
    \eta_C = 1-\frac{T_c}{T_h}.
\end{align}
The Curzon-Ahlborn efficiency 
\begin{align}
    \eta_{CA} = 1-\sqrt{\frac{T_c}{T_h}} \approx  \frac{\eta_C}{2} + \frac{\eta_C^2}{8} + \dots \label{eq:CA}
\end{align}
provides a bound on the efficiency at maximum power \cite{Curzon1975,Esposito2009}.
Unlike the Carnot bound it can be violated, e.g. for a single level QD in the sequential tunneling limit \cite{VandenBroeck2005}.

\subsection{Thermodynamic uncertainty relations}
With the help of the current noise from counting statistics we can investigate the TUR \cite{Barato2015,Pietzonka2018,Seifert2018,Gingrich2016,Agarwalla2018,Liu2019}
\begin{align}
    \frac{S}{I^2}\sigma \geq 2.
\end{align}
In the heat engine case the entropy production reads $\sigma = P\frac{1}{T_c}\frac{\eta_C-\eta}{\eta}$ \cite{Whitney2015, Liu2019} and inserting this results in 
\begin{align}
    \frac{S}{I}V\frac{1}{T_c}\frac{\eta_C-\eta}{\eta} \geq 2. \label{eq:TUR}
\end{align}
We can rearrange Eq. \eqref{eq:TUR} to get bounds on the individual quantities given the other two, e.g. for the noise 
\begin{align}
    S \geq 2 I \frac{T_c}{V} \frac{\eta}{\eta_C-\eta}, \label{eq:TURS}
\end{align}
or the efficiency
\begin{align}
    \eta \leq \frac{\eta_C}{\frac{2IT_c}{SV}+1} \label{eq:TURn}.
\end{align}

\section{Results}

\begin{figure*}
    \centering
    \includegraphics[width=\textwidth]{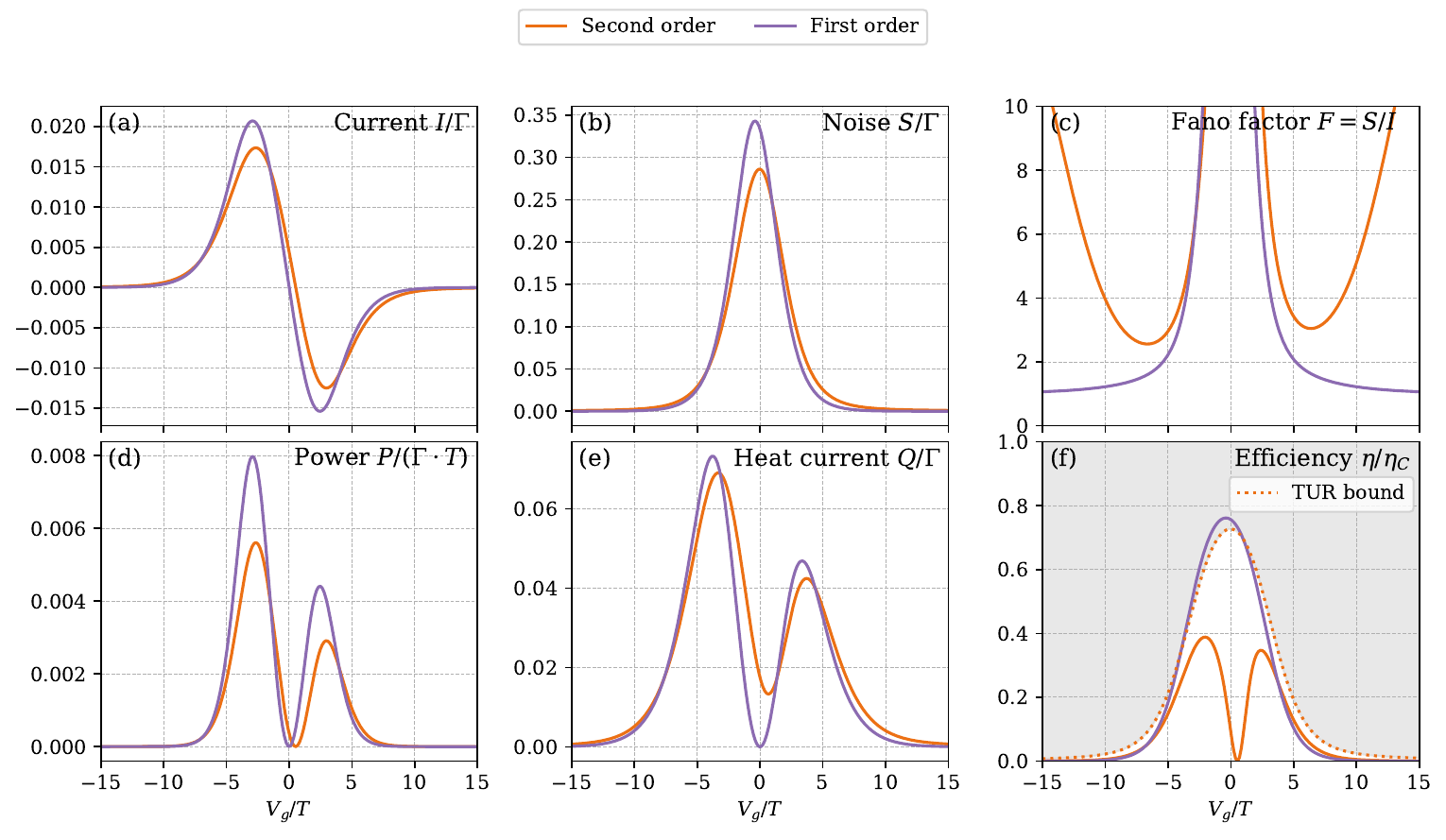}
    \caption{
    (a) Current, 
    (b) noise,
    (c) Fano factor,
    (d) power,
    (e) heat current and 
    (f) efficiency 
    as a function of $V_g$ in the setup in \figref{fig:1}(c).
    The resistance $R\approx 74.43$ is chosen to maximize the power.
    All other parameters are the same as in \figref{fig:1}.
    All figures show second order (orange) and first order (purple) results. 
    The orange dotted line in the efficiency (f) indicates the bound given by the TUR \eqref{eq:TURn} for the second order, where the shaded areas are indicate TUR violations.
    Note, that we choose the optimal load of the second order approach for both cases, since the difference is small.
    }
    \label{fig:2}
\end{figure*}

Figure \ref{fig:2} shows different transport quantities along a gate sweep for the setup shown in \figref{fig:1}(c), i.e. the QD coupled in a circuit with an external load resistance $R$ and no external bias voltage.
In each case we compare the result from first and second order perturbation theory in the tunnel couplings, referred to as first and second order respectively.

The current, \figref{fig:2}(a), is positive for gate voltages left of the charge degeneracy point.
This corresponds to the situation sketched in \figref{fig:1}(a), where the QD level sits above the lead chemical potentials (charge $0$ sector).
Note that the built up voltage is negative here, i.e. $\mu_h<\mu_c$, and positive current in our convention means electrons flow from the hot lead into the QD.
For large negative gate voltages the level is far away from the bias window and the current is small.
Getting closer to the bias window the current increases, due to the increasing difference in the Fermi functions of the leads.
At the charge degeneracy point the chemical potentials line up with the QD level and the current has to be zero.
To the right of the degeneracy point (charge $1$ sector) the occupation in the cold lead is larger and the current is inverted compared to the charge $0$ sector.
The asymmetry in the height of the peak between the charge sectors results from a combination of spin degeneracy and Coulomb interaction, see Appendix A of Ref. \cite{Josefsson2019}.
Note that at the 1-2 charge degeneracy point the picture would be flipped.
When comparing first and second order in \figref{fig:2}(a) we observe the known shift of the current zero due to second order contributions.
This is due to a re-normalization of the level energies in second order \cite{Koenig1998,Kubala2006,Leijnse2008} and is also visible in the other plots.

The noise $S$ in \figref{fig:2}(b) is a positive quantity.
It shows one peak centered at the charge degeneracy point and looks qualitatively similar in first and second order.
The second order noise approaches zero slower than the first order noise.

The effect of the second order tunneling becomes more clear in the Fano factor $F=S/I$ in \figref{fig:2}(c).
Towards the charge degeneracy it has to diverge, since the current in the denominator becomes 0, while the noise stays finite.
In first order the Fano factor converges to a fixed value in the blockaded regime, i.e. the current and noise converge towards zero at the same rate.
In contrast, the second order Fano factor does not converge when going into the blockaded regime, because in second order the current goes to zero faster than the noise.
This can be explained by additional transport mechanisms in second order that can contribute to the noise, but not necessarily to the current \cite{DeFranceschi2001,Thielmann2005}.

The power in \figref{fig:2}(d) is developed in the load.
It has to be zero at the current inversion point, with peaks on either side of it.
The asymmetry in the current results in an asymmetry in the power.
Here also the shift of the current inversion point in second order mentioned earlier becomes clear.

The heat current $Q_h$ at the hot lead, \figref{fig:2}(e), is coupled to the electron current and thus has a dip at the current inversion point, peaks on either side and vanishes for larger $|V_g|$.
In second order, $Q_h$ does not go to zero at the current inversion point.
This is because the tight coupling between particle and heat current gets broken by second order tunneling \cite{Josefsson2018,Josefsson2019}.
This leads to a striking difference in the efficiency $\eta=P/Q_h$, shown in \figref{fig:2}(f), which reduces to zero at the current inversion point in second order, but not in first order.
The orange dotted line in \figref{fig:2}(f) indicates the bound given by the TUR \eqref{eq:TURn} for the second order curve.
Note that because the bound includes $I$ and $S$, it is not completely identical in first and second order.
We omit the first order bound, since at this scale it cannot be distinguished from the efficiency.
The bound is not saturated in second order.
Notably, the TUR bound is smaller than the Carnot efficiency, i.e. given the current and noise of a heat engine and assuming the TURs to hold, we can find a tighter bound than Carnot.
TURs can in principle be violated in second order \cite{Liu2019,Agarwalla2018}.
However, we find that this does not happen in our system, at least not in the parameter regime we investigate.

\begin{figure*}
    \centering
    \includegraphics[width=\textwidth]{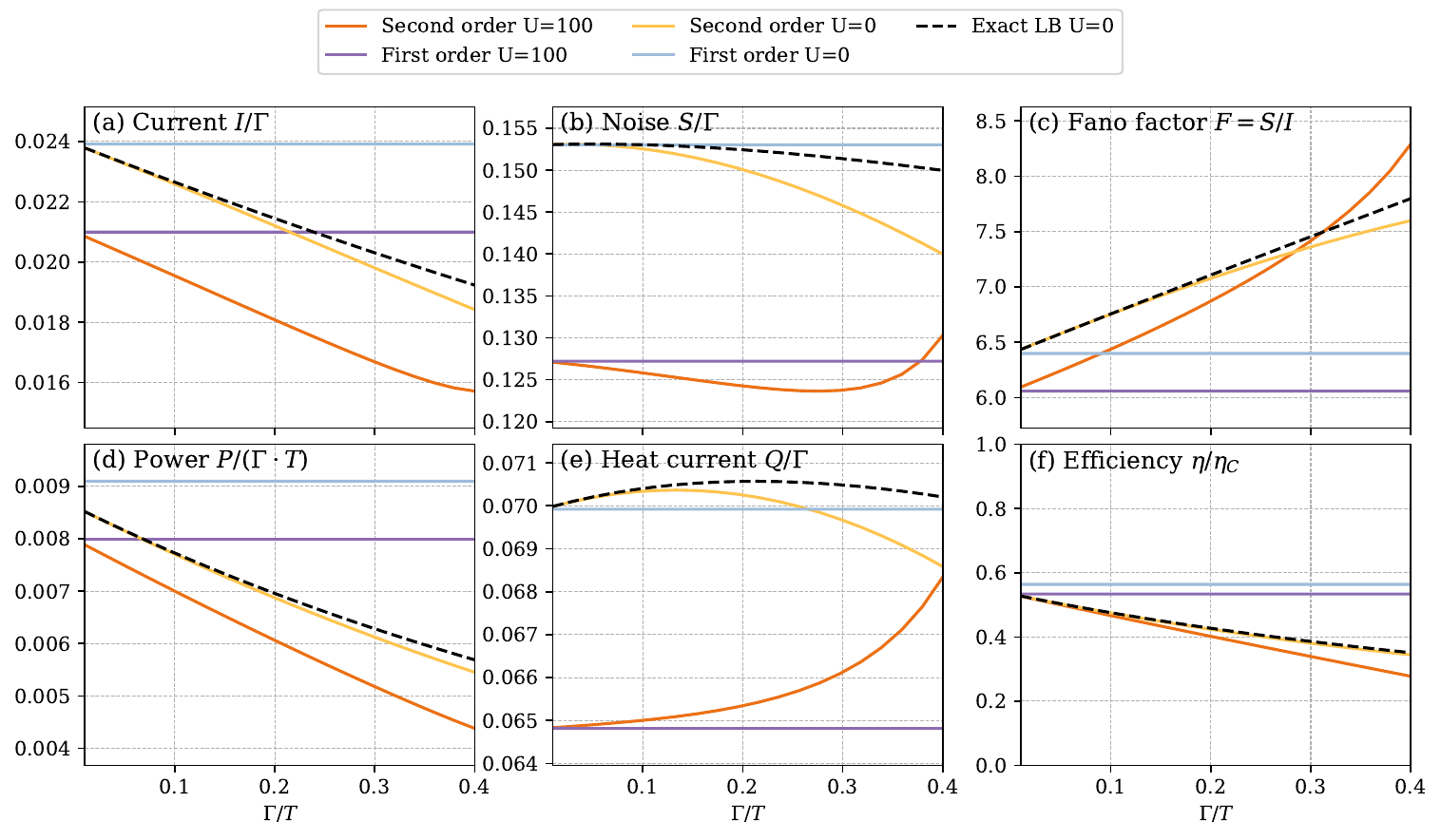}
    \caption{
    (a) Current, 
    (b) noise,
    (c) Fano factor,
    (d) power,
    (e) heat current and 
    (f) efficiency 
    at maximum power plotted as a function of the tunnel coupling to the leads for different approaches and different values of $U$ (see top of Figure).
    Other parameters are the same as in \figref{fig:1}.
    }
    \label{fig:3}
\end{figure*}

In \figref{fig:3} we show the same quantities as in \figref{fig:2}, but at the maximum power point varying the tunnel couplings to the leads.
We also show different Coulomb interaction strengths, $U=100$ and $U=0$, and compare to scattering theory for $U=0$.
In first order the maximum power point does not change with $\Gamma$.
Because $\Gamma$ is simply a pre-factor in the first order rate equation the current, noise, power and heat current for the first order approach are linear in $\Gamma$.
Note the scaling with $\Gamma$ that leads to all these quantities being constant in first order.
Deviations from that behavior are a related to higher order tunneling.
For second order and non-interacting scattering theory the maximum power point moves in gate-bias space when sweeping $\Gamma$.

Figure \ref{fig:3}(a) shows the currents scaled by $\Gamma$.
Without interactions, the second order current follows the scattering theory result closely.
Coulomb interactions reduce the current for both first and second order.
Towards $\Gamma \approx 0.4$ the second order current changes curvature indicating breakdown of perturbation theory.
Similar behavior can be seen in the noise in \figref{fig:3}(b).
The deviation between scattering theory and second order noise can be explained by the maximum power point shifting more in second order, compared to scattering theory (see \figref{fig:maxpp} in Appendix \ref{app:maxpp}).
The noise at the maximum power point is sensitive to changes in gate position (compare Figs. \ref{fig:2}(b) and (d)) leading to large deviations.

Figure \ref{fig:3}(c) shows the Fano factor, which increases with $\Gamma$ in the non-interacting second order and scattering theory results.
For bigger tunnel couplings the deviations become larger.
In first order, Coulomb interaction reduces the Fano factor.
For small $\Gamma$ this is also the case for second order, but for larger tunnel coupling the interactions lead to a crossover of the two curves.

Figure \ref{fig:3}(d) shows the power.
The second order curve without Coulomb interactions follows the scattering theory results well.
Coulomb interactions reduce the power in both first and second order.
In the heat current, \figref{fig:3}(e), second order without interactions and scattering theory results qualitatively agree.
In both first and second order, Coulomb interactions lead to a reduced heat current.
The efficiency, \figref{fig:3}(f), for second order and scattering theory both show a decrease in efficiency with increasing $\Gamma$.

\begin{figure*}
    \centering
    \includegraphics[width=\textwidth]{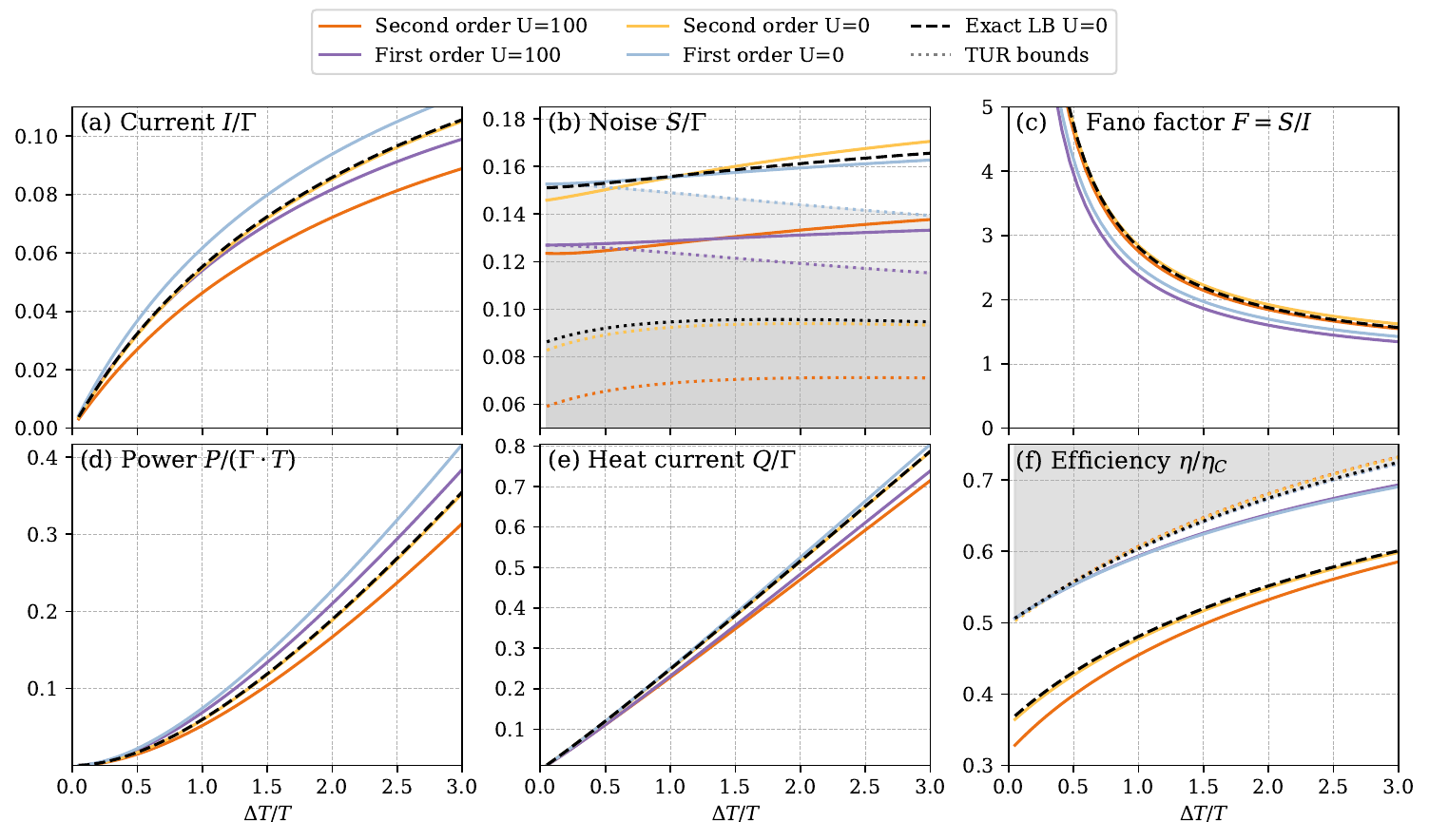}
    \caption{
    (a) Current, 
    (b) noise,
    (c) Fano factor,
    (d) power,
    (e) heat current and 
    (f) efficiency 
    at maximum power plotted as a function of the temperature difference $\Delta T$ between the leads for different approaches and different values of $U$ (see top of Figure).
    Other parameters are the same as in \figref{fig:1}.
    The dotted lines in the noise (b) and efficiency (f) indicate the bound given by the TUR, Eq. \eqref{eq:TUR}, for the respective curves, where the shaded areas are forbidden.
    }
    \label{fig:4}
\end{figure*}

In \figref{fig:4} we vary the temperature difference $\Delta T$ by increasing the temperature of the hot lead, while keeping the temperature of the cold lead fixed.
As before the position of the maximum power point will shift in gate with increasing temperature difference, also requiring a different optimal load.
Here it will shift for both first and second order, as well as for scattering theory, see \figref{fig:maxppdt} in Appendix \ref{app:maxpp}.

The current in \figref{fig:4}(a) increases with the temperature difference.
For $\Delta T \gtrsim 0.5$ this increase is clearly smaller than linear.
The noise in \figref{fig:4}(b) increases with the temperature difference.
An increased temperature of one of the leads means the average temperature of the system is higher.
That is expected to result in more thermal noise.
Similarly to the current, the noise is reduced by the interactions.
The bounds given by the TUR in Eq. \eqref{eq:TURS} are given as dotted lines in the corresponding color.
For small $\Delta T$ the first order results with Coulomb interaction seem to saturate the bound, while the other results are all far away from saturation.
Increasing the temperature difference increases the noise in all cases and pushes the noise further away from the bound.

For decreasing temperature difference the maximum power point converges to the charge degeneracy point, where the current is zero, while the noise is finite (and the values is given by the equilibrium fluctuation dissipation theorem as $S=-2\bar{T}\partial I / \partial\Delta\mu|_{\Delta\mu =0}$).
Thus, the Fano factor in \figref{fig:4}(c) diverges for small temperature differences.

The power in \figref{fig:4}(d) increases with increasing temperature difference, since at larger $\Delta T$ the QD supports larger currents.
The heat current at the hot lead increases with the temperature difference in all cases, see \figref{fig:4}(e).
The efficiencies in \figref{fig:4}(f) are qualitative similar, where the first order efficiencies are higher than the other results and follow the Curzon-Ahlborn efficiency in Eq. \eqref{eq:CA} (not shown here).
Again we indicate the TUR bound from Eq. \eqref{eq:TURn} by dotted lines.
These bounds are consistently smaller than the Carnot efficiency.

\section{Conclusion}

In this work, we have theoretically studied a QD operated as a thermoelectric heat engine.
In particular, we used a second order master equation approach to calculate current, noise and heat current, as well as quantities derived from these.
For the QD in a heat engine configuration we investigate a gate sweep when coupled to a load resistance, as well as sweeping the tunnel couplings and temperature difference at maximum power output.
We compare results from first and second order perturbation theory in the tunneling rates $\Gamma$, and in  the non-interacting case we additionally compare with exact scattering theory.
We observe that in the heat engine regime the second order processes can reduce the noise.
However, at the same time the current is suppressed and the Fano factor typically increases compared to pure sequential tunneling approaches.
Coulomb interactions reduce the noise and lower the Fano factor, at least in the parameter range we investigated.

Since the TURs contain the Fano factor, second order tunneling pushes the heat engine further away from saturating the TUR.
While interactions reduce the Fano factor, the reduction of the heat current and thus efficiency means that the TUR is also further from being saturated for significant Coulomb interaction strength.
Our results show no TUR violations when operating a QD as a thermoelectric engine at maximum power.
In that sense they provide a bound on the efficiency of the engine (given current and noise) that is lower than the fundamental Carnot efficiency.

\begin{acknowledgments}
We thank Konstantin Nestmann for valuable input on the transport theory, and Patrick Potts, Peter Samuelsson and Adam Burke for stimulating discussions about TURs.
We acknowledge funding from the Swedish Research Council under Grant Agreement No. 2020-03412 and from NanoLund.
\end{acknowledgments}


%

\appendix

\section{Landauer-B\"uttiker scattering theory} \label{app:lb}

Here we present the resulting equations for the current $I$, noise $S$ and heat currents $Q_r$ for a single electronic level at $\varepsilon$ coupled to leads at temperatures $T_r$ and chemical potentials $\mu_r$.
For a derivation see e.g. chapter 2.3-2.4 in Ref. \cite{Blanter2000}.
The current and noise read
\begin{align}
I = \frac{1}{2\pi}\int dE \lbrace T(E)[f_h(E)-f_c(E)] \rbrace,
\end{align}
\begin{widetext}
\begin{align}
S = \frac{1}{2\pi}\int dE \lbrace T(E)[f_h(E)(1-f_h(E))+f_c(E)(1-f_c(E))] + T(E)[1-T(E)](f_h(E)-f_c(E))^2 \rbrace.
\end{align}
\end{widetext}
The transmission $T(E)$ for a single resonant level at $\varepsilon$ given the tunnel rates $\Gamma_{h/c}$ is given by a Lorentzian
\begin{align}
T(E) = \frac{\Gamma_h \Gamma_c}{(E-\varepsilon)^2 + (\Gamma_h + \Gamma_c)^2/4}.
\end{align}
To find the heat current we first use the energy current
\begin{align}
J_E = \frac{1}{2\pi}\int dE \lbrace E \cdot T(E)[f_h(E)-f_c(E)] \rbrace.
\end{align}
As in the main text the heat current at lead $r$ then reads
\begin{align}
Q_r = J_E - \mu_r I.
\end{align}
Note that the result for the non-interacting Anderson dot is simply the results of the single levels times two.

\section{Postition of the max power point}\label{app:maxpp}

The area in gate/bias space where the QD operates as a heat engine, i.e. has positive power output, changes when changing parameters such as the coupling to the leads or the temperature difference between the leads.
We are interested in comparing how the transport quantities and heat engine characteristics at the point of maximum power output evolves when sweeping parameters.

The maximum power point $(V_{g,\textrm{max}},V_{b,\textrm{max}})$ in gate-bias space can be found numerically by optimizing the power $P=-IV_b$ with an applied external bias voltage $V_b$.
The resulting point in \figref{fig:1}(a) is marked with a red cross.
This bias voltage corresponds to choosing a load resistance $R=V_{b,\textrm{max}}/I(V_{g,\textrm{max}},V_{b,\textrm{max}})$ and tuning to the gate voltage $V_{g,\textrm{max}}$.
Finding the voltages $V$ that develop over the resistor for each gate voltage then results in the red dashed line in \figref{fig:1}(b).

\begin{figure}
    \centering
    \includegraphics[width=\columnwidth]{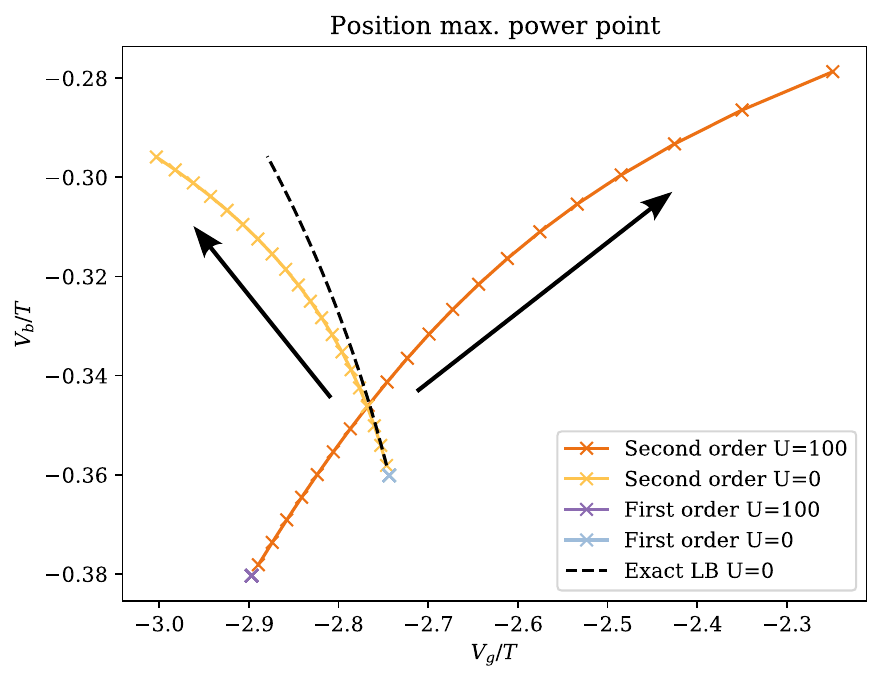}
    \caption{
    Position of the max power point in gate/bias space depending on the tunnel couplings $0.01 \leq \Gamma \leq 0.4$ (arrows indicate increasing $\Gamma$).
    }
    \label{fig:maxpp}
\end{figure}

Figure \ref{fig:maxpp} shows the position of the maximum power point for increasing $\Gamma$.
Note that the maximum power point for the first order does not shift at all.
This is because in first order $\Gamma$ is only a pre-factor, scaling the current and hence the power.
In second order and scattering theory the maximum moves towards lower bias, in the direction of the arrows.
In gate the Coulomb interaction determines the direction of the shift.

\begin{figure}
    \centering
    \includegraphics[width=\columnwidth]{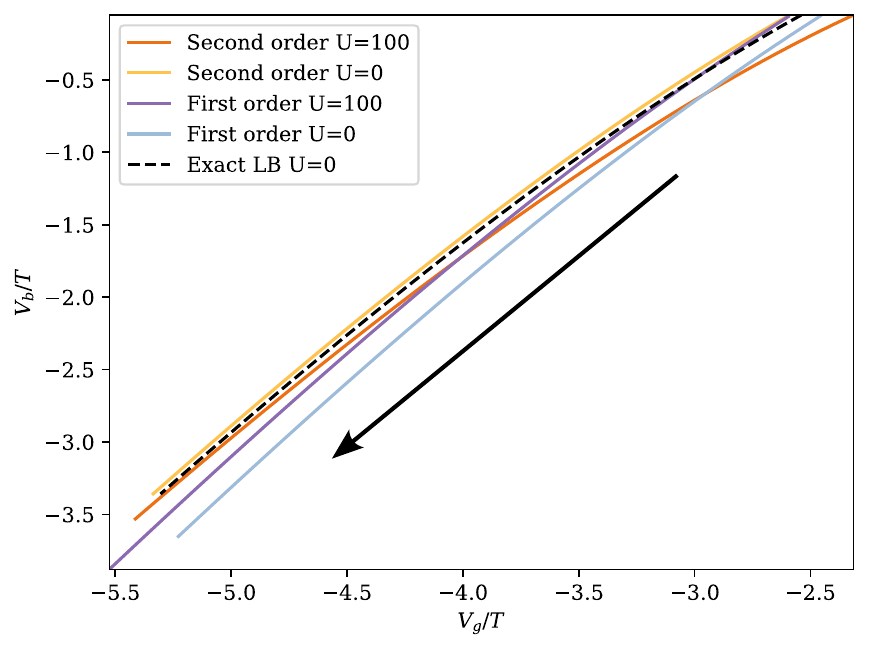}
    \caption{
    Position of the max power point in gate/bias space depending on the temperature difference $0.05 \leq \Delta T \leq 3.0$ (arrow indicates increasing $\Delta T$).
    }
    \label{fig:maxppdt}
\end{figure}

For a sweep in temperature difference the max power point moves towards larger gate and bias voltages for all approaches in a similar fashion, see \figref{fig:maxppdt}.

\end{document}